\documentclass[12pt]{iopart}

\usepackage{graphicx}

\begin{document}

\title{Invisibility in $\mathcal{PT}$-symmetric complex crystals}

\author{Stefano Longhi}

\address{Dipartimento di Fisica,
Politecnico di Milano, Piazza L. da Vinci 32, I-20133 Milano, Italy}
\ead{longhi@fisi.polimi.it}
\begin{abstract}
Bragg scattering in sinusoidal $\mathcal{PT}$-symmetric complex crystals of finite thickness is
theoretically investigated by the derivation of exact analytical expressions for reflection and transmission coefficients in terms of 
modified Bessel functions of first kind. The analytical results indicate that unidirectional invisibility, recently predicted for such crystals  by coupled-mode theory  [Z. Lin {\em et al.}, Phys. Rev. Lett. {\bf 106}, 213901 (2011)], 
breaks down for crystals containing a large number of unit cells. In particular, for a given modulation depth in a shallow sinusoidal potential, three regimes are encountered as the crystal thickness is increased. At short lengths  the crystal is reflectionless and invisible when probed from one side (unidirectional invisibility), whereas at intermediate lengths the crystal remains reflectionless but not invisible; for longer crystals both unidirectional reflectionless and invisibility properties are broken.
\end{abstract}


\maketitle

\section{Introduction}
 Physical phenomena described by reduced or effective non-Hermitian Hamiltonians are 
often encountered  in a wide class of quantum or classical systems, for example in nuclear or condensed-matter physics of open systems \cite{NH1,NH2,NH3}  or in optical systems
 in presence of optical gain or losses \cite{NH4}. Among non-Hermitian Hamiltonians, great interest has been devoted in the past two decades to study the properties of  parity-time ($\mathcal{PT}$) invariant Hamiltonians, which possess a real-valued energy spectrum below a  symmetry-breaking point in spite of non-Hermiticity. Such a class of non-Hermitian Hamiltonians has been originally introduced by Carl Bender in the framework of non-Hermitian extensions of quantum mechanics and quantum field theories \cite{Bender1,Bender2,Bender3}, and found recently 
an increasing interest since the proposal of physical systems described by  $\mathcal{PT}$-symmetric Hamiltonians, including optical  \cite{O1,O2,O3,O4,O5,O6,O7,O8,Science,refe1,refe2,refe3} and electronic \cite{electro} systems.
 \par
Complex periodic potentials \cite{C1,C2,C3,C4,C4bis,C5,C6,C6tris,C6bis,C7} realize a kind of synthetic complex crystals, which show rather unusual   scattering and transport properties as compared to ordinary crystals.  Complex crystals
have been investigated in different areas of physics, ranging
from matter waves \cite{C4bis,C5,C6,C6tris,C6bis} to optics \cite{O3}.  
In optics, a complex crystal with $\mathcal{PT}$ invariance is realized by 
introduction of index and balanced gain/loss modulations in a dielectric medium \cite{O3}. 
Complex crystals can be also realized in atom optics experiments exploiting
the interaction of near resonant light with an open two-level
system. Scattering of matter waves from purely absorbing optical lattices was reported in a few earlier experiments Refs.\cite{C4bis,C5,C6bis}. As noticed by Berry \cite{Berry},  although in such experiments on matter waves $\mathcal{PT}$ symmetry is not strictly realized and the complex potentials are purely absorptive, the analysis is essentially the
same, since the mean loss simply represents an overall exponential decay of the wave. 
From the theoretical side, Bragg scattering, diffraction and transport properties have been extensively investigated for  sinusoidal $\mathcal{PT}$-symmetric  complex crystals \cite{O3,C1,C2,L1,L2,L3,L4,L5}, revealing some interesting properties such as  violation of the Friedel's law of Bragg
scattering \cite{C5,C6,L1}, double refraction and nonreciprocal diffraction \cite{O3},
and unidirectional Bloch oscillations \cite{L4}.  In particular, in a recent work \cite{L5}  it was predicted that a sinusoidal $\mathcal{PT}$-symmetric sinusoidal crystal of finite length near the spontaneous $\mathcal{PT}$-symmetry breaking point
can act as a unidirectional invisible medium, i.e. the crystal is  almost reflectionless when probed from one side, and transmission occurs as if the crystal were absent.  Such an unidirectional invisibility of $\mathcal{PT}$-symmetric Bragg scatters near the symmetry-breaking point was previously predicted to occur in Ref.\cite{cg} for waveguide Bragg gratings which
combine matched periodic modulations of refractive index and loss/gain
yielding asymmetrical mode coupling. In these previous studies \cite{L5,cg}, invisibility was explained on the basis of  a coupled-mode theory  describing Bragg scattering and coupling of counter-propagating waves in the crystal, which is rather common in the optical context \cite{Sipe,Poladian}. Such an analysis predicts that, for a shallow grating near the $\mathcal{PT}$ symmetry breaking point, the sinusoidal crystal appears to be invisible when probed from one side {\it independently} of the crystal length. \par 
In this work  we re-consider the scattering properties of  the sinusoidal $\mathcal{PT}$-symmetric potential and derive {\it exact} analytical expressions for reflection and transmission coefficients. The analysis shows that application of the coupled-mode theory in the standard form fails to predict the correct scattering properties in case of long crystals. In particular, as at short lengths  the crystal is reflectionless and invisible when probed from one side (according to previous studies \cite{L5,cg}), at intermediate lengths the crystal remains reflectionless but not invisible. At even longer crystal lengths, both unidirectional reflectionless and invisibility properties are broken.

\section{Bragg scattering in $\mathcal{PT}$-symmetric sinusoidal potentials: general aspects and extended coupled-mode theory}

\subsection{The model}

Let us consider the stationary Schr\"{o}dinger equation for a quantum particle in a locally periodic and complex potential V(x), which in dimensionless form reads
\begin{equation}
\hat{H} \psi \equiv -\frac{d^2 \psi}{dx^2}-V(x) \psi=E \psi
\end{equation}
where $E$ is the energy of the incident particle and $V(x)$ is the complex scattering potential with period $\Lambda$, which is nonvanishing in the interval $0<x<L$. The crystal length $L$ is assumed to be an integer multiple of the lattice period $\Lambda$, i.e. $L=N \Lambda$, where $N$ is the number of unit cells in the crystal. As mentioned in the introduction, Eq.(1) describes Bragg scattering of matter waves from a complex potential in the non-interacting regime, which applies e.g. to a dilute  cold atomic beam (see, for instance, \cite{C6bis}). In this case, the complex potential arises from the interaction of near resonant light with an open two-level
system, and it is generally absorptive. In this work we will mainly focus our attention to the $\mathcal{PT}$-symmetric sinusoidal potential, assuming
\begin{equation}
V(x)=V_0 \left[ \cos \left( 2 \pi x / \Lambda \right)+i \sigma \sin \left( 2 \pi x / \Lambda\right) \right]
\end{equation}
for $0<x<L$, and $V(x)=0$ for $x<0$ and $x>L$, 
where $V_0$ is the lattice amplitude and $\sigma \geq 0$ measures the strength of the non-Hermitian part of the potential. The spectral properties of the $\mathcal{PT}$-symmetric sinusoidal potential (3) have been investigated in Refs.\cite{O3,C2,C7,L1,L4}. For the infinitely-extended crystal, the energy spectrum remains real-valued for $\sigma \leq 1$, and breaking of the $\mathcal{PT}$ phase is attained at $\sigma=\sigma_c=1$ \cite{O3,C2,C7,L1,L4}. Here we consider Bragg scattering of incoming waves with momentum $p$ close to the Bragg value $\pi/ \Lambda$, i.e. with energy $E=p^2$ close to $(\pi/\Lambda)^2$, and typically will assume a modulation $V_0$ of the potential much smaller than the energy $E$. \\
It should be noted that Bragg scattering of optical waves in one-dimensional Bragg grating structures, considered in Refs.\cite{L5,cg}, is basically analogous to Bragg scattering of matter waves in the framework of Eq.(1). In fact, the electric field amplitude $\mathcal{E}(x)$ of an optical wave at frequency $\omega$ that propagates along a dielectric medium with a spatially-dependent relative dielectric constant $\epsilon(x)=n_0^2[1+\Delta \epsilon (x)]$, where $n_0$ is the refractive index of the lossless medium and $\Delta \epsilon(x+\Lambda)=\Delta \epsilon(x)$ accounts for the 
index and gain/loss modulation, satisfies the scalar Helmholtz equation, which can be written in the form
\begin{equation}
-\frac{d^2 \mathcal{E}}{dx^2}-E \Delta \epsilon(x)  \mathcal{E}=E \mathcal{E},
\end{equation}
where we have set $E=k^2$ and $k= n_0 \omega/c_0$. Note that Eq.(3) formally reduces to the Schr\"{o}dinger equation provided that the following formal substitutions 
\begin{equation}
\mathcal{E} \rightarrow \psi \; ,  \;\;\; k \rightarrow p 
\end{equation}
are made, with a complex scattering potential $V(x)$ related to the modulation of the dielectric constant $\Delta \epsilon(x)$ by the simple relation
\begin{equation}
 V(x)=E \Delta \epsilon(x). 
\end{equation}
Hence, the only difference between scattering of matter waves in complex optical potentials and light waves in complex Bragg gratings is that, in the latter case, the complex scattering potential $V(x)=E\Delta \epsilon(x)$ in the equivalent Schr\"{o}dinger equation depends
on the energy $E$  of the incidence particle. However, for shallow gratings Bragg scattering occurs solely for optical fields with frequencies $\omega$ very close to the Bragg frequency $\omega_B=c_0 \pi/(n_0 \Lambda)$ (see, e.g. \cite{cg}), and thus one can safely assume $V(x) \simeq (\pi/ \Lambda)^2 \Delta \epsilon(x)$, leaving out the dependence of the scattering potential from the energy. In the following, we will mainly focus our analysis to the determination of the reflection and transmission coefficients for scattering of matter waves in the framework of Eq.(1), however similar results hold {\it mutatis mutandis} for reflection and transmission of optical waves in Bragg grating structures. 

\subsection{Scattering states, spectrum, and reflection/transmission coefficients}
Since $V(x)=0$ for $x<0$ and $x>L$, the continuous spectrum of the Hamiltonian $\hat{H}$ is the  semi-infinite real axis of energies $E=p^2 \geq 0$, and the corresponding eigenfunctions are the scattered states, defined by the relations
\begin{equation}
\psi(x)= \left\{ 
\begin{array}{c}
\alpha_1 \exp(ipx)+\beta_1 \exp(-ipx) \; \; x \leq 0 \\
\alpha_2 \exp[ip(x-L)+\beta_2 \exp[-ip(x-L)] \; \; x \geq L \\
\end{array}
\right.
\end{equation}
where $p \geq 0$ is the momentum and $(\alpha_1,\beta_1)$, $(\alpha_2,\beta_2)$ are the amplitudes of forward and backward propagating waves on the left ($x<0$) and on the right ($x>L$) sides of the crystal, respectively. Such amplitudes are related by the algebraic equation (see, for instance, \cite{O6})
\begin{equation}
\left(
\begin{array}{c}
\alpha_2 \\
\beta_2
\end{array}
\right)=\mathcal{M}(p) \left(
\begin{array}{c}
\alpha_1 \\
\beta_1
\end{array}
\right)
\end{equation}
where the $2 \times 2$ transfer matrix $\mathcal{M}(p)$ is unimodular, i.e. ${\rm det} \mathcal{M}=\mathcal{M}_{22} \mathcal{M}_{11}-\mathcal{M}_{12}\mathcal{M}_{21}=1$. For a $\mathcal{PT}$-symmetric potential,  the further relation $\mathcal{M}_{22}(p)=\mathcal{M}_{11}^*(p^*)$ holds.
 The transmission ($t$) and reflection
($r$) coefficients for left ($l$) and right ($r$) side incidence are related to the coefficients of the transfer matrix by the usual relations (see, for instance, \cite{O6})
 \begin{equation}
 t^{(l)}= \frac{1}{\mathcal{M}_{22}} , \;  \;  t^{(r)}=t^{(l)} \equiv t , \;\;\;\; 
 r^{(l)}=-\frac{\mathcal{M}_{21}}{\mathcal{M}_{22}}, \; \;   r^{(r)}=\frac{\mathcal{M}_{12}}{\mathcal{M}_{22}}. 
 \end{equation} 
Note that the transmission coefficient does not depend on the incidence side like in an ordinary crystal, whereas generally one has  $|r^{(l)}| \neq |r^{(r)}|$, i.e. in reflection a complex crystal behaves differently for left and right incidence. This is  a rather general result of wave scattering from a complex potential barrier which was previously discussed e.g. in \cite{ahmed,ventura}. If we indicate by $\mathcal{Z}(p)$ the fundamental matrix of Eq.(1) from $x=0$ to $z=L$ which relates the values of $\psi(x)$ and $(d \psi/dx)$ at the planes $x=0$ and $x=L$, i.e.
\begin{equation}
\left(
\begin{array}{c}
\psi(L) \\
(d \psi / dx) (L)
\end{array}
\right)=
\mathcal{Z}(p) \left(
\begin{array}{c}
\psi(0) \\
(d \psi / dx) (0)
\end{array}
\right)
\end{equation}
it can be readily shown that the transfer matrix $\mathcal{M}$ can be calculated as
\begin{equation}
\mathcal{M}(p)=\mathcal{T}^{-1}(p)  \mathcal{Z}(p)  \mathcal{T}(p)
\end{equation}
where we have set
\begin{equation}
\mathcal{T}(p) = \left(
\begin{array}{cc}
1 & 1 \\
ip & -ip
\end{array}
\right).
\end{equation}
From a numerical viewpoint, the fundamental matrix $ \mathcal{Z}(p)$ can be computed as follows. Let us cut the crystal into a sequence of $N_0$ thin slices, of thickness $\Delta x=L/N_0$,  and let us indicate by $ \mathcal{Z}_k (p)$ the fundamental matrix associated to the propagation at the $k$-th slice ($k=1,2,...,N_0$), i.e. from $x_k=(k-1)\Delta x$ to $x_{k+1}=k \Delta x$. Then the fundamental matrix $ \mathcal{Z}(p)$
can be calculated as the ordered product
\begin{equation}
\mathcal{Z}(p)=\mathcal{Z}_N(p) \times  \mathcal{Z}_{N-1}(p) \times ... \times \mathcal{Z}_2(p) \times \mathcal{Z}_1(p)
\end{equation}
If $\Delta x$ is much smaller than $\Lambda$, the potential $V(x)$ is almost constant in the interval $(x_{k},x_{k+1})$, and thus  $ \mathcal{Z}_k (p)$  can be approximated as
\begin{equation}
\mathcal{Z}_k (p) \simeq \left( 
\begin{array}{cc}
\cos(\lambda_k \Delta x) & (1/\lambda_k) \sin(\lambda_k \Delta x) \\
- \lambda_k \sin(\lambda_k \Delta x) & \cos(\lambda_k \Delta x)
\end{array}
\right)
\end{equation}
where we have set
\begin{equation}
\lambda_k=\sqrt{p^2+V(x_k)}.
\end{equation}
Note that, because of the periodicity of the crystal, one can limit to compute the fundamental matrix for the unit cell, $\mathcal{Z}^{(cell)}(p)$, i.e. from $x=0$ to $x= \Lambda$. The fundamental matrix of the crystal is then given by $\mathcal{Z}(p)=\mathcal{Z}^{(cell) N}(p) $, which can be computed using the relation \cite{note} $\mathcal{Z}(p)=\mathcal{Z}^{(cell)}(p) \mathcal{U}_{N_1}-\mathcal{I} \mathcal{U}_{N-2} $, where $\mathcal{I}$ is the $2 \times 2 $ identity matrix, $\mathcal{U}_N=\sin[(N+1) \theta]  / \sin \theta$, and the complex angle $\theta$ is defined by $\cos \theta=(1/2) {\rm Tr} (\mathcal{Z}^{(cell)})$.

Besides scattering states, the Hamiltonian $\hat{H}$ can possess bound states belonging to the point spectrum, which are determined by the zeros of $\mathcal{M}_{22}(p)$ (i.e. the poles of the transmission coefficient $t$) in the half complex plane ${\rm Im}(p)>0$. The zeros of $\mathcal{M}_{22}(p)$ in the half complex plane ${\rm Im}(p)<0$
 are resonance states.  The onset of $\mathcal{PT}$ symmetry breaking is detected by the appearance of a  divergence (for some real value $p=p_0 \neq 0$ of the momentum) in the transmission coefficient as the non-Hermiticity parameter $\sigma$ is increased from zero. In fact, for $\sigma=0$ the transmission and reflection coefficients are bounded from above and the zeros of $\mathcal{M}_{22}(p)$ lie in the lower half complex plane  ${\rm Im}(p)<0$, i.e. they are resonances.  As $\sigma$ in increased, the resonances move toward the real axis ${\rm Im}(p)=0$, until at a critical value $\sigma=\sigma_c$ a resonance crosses the real axis, say at $p=p_0 \neq 0$ (real). Correspondingly, the transmission coefficient $t(p)$ diverges at $p=p_0$.  Just above $\sigma_c$, a bound state with complex energy $E=(p_0+i0^+)^2$  thus appears, which is the signature of $\mathcal{PT}$ symmetry breaking. 
 For a sinusoidal crystal of {\it finite} length, the transmission and reflection coefficients at $\sigma=1$ are bounded, and numerical calculations of the transfer matrix $\mathcal{M}$ (using the procedure outlined above) indicate that symmetry breaking is attained at a value $\sigma_c$ {\em  larger} than one, with $\sigma_c \rightarrow 1^+$ as $L \rightarrow \infty$.

\subsection{Coupled-mode theory}

For a rather general class of $\mathcal{PT}$-symmetric complex potentials $V(x)$ describing Bragg scattering in {\it shallow} lattices,  approximate expressions for the reflection and transmission coefficients can be derived by an asymptotic analysis of Eq.(1).  In the optical context, such an analysis is generally referred to as the coupled-mode theory of Bragg scattering, which is known to  provide accurate description of transmission and reflection coefficients for index-modulated shallow gratings (see, for instance,  \cite{Sipe, Poladian}). 
Such an analysis applies to Bragg scattering of particles with momentum $p$ close to $\pi / \Lambda$ (first-order Bragg scattering) provided  that  the particle energy  $E  \simeq (\pi/ \Lambda)^2$ is much larger than the characteristic modulation depth $V_0$ of the complex crystal. For the sinusoidal crystal defined by Eq.(2), this means that the parameter $\alpha=\Lambda^2 V_0/ \pi^2$ should be much smaller than one.  In Ref.\cite{L5,cg}, it was shown that application of coupled-mode theory to the sinusoidal $\mathcal{PT}$-symmetric crystal
at $\sigma=1$ gives the following expressions for the transmission and reflection coefficients
\begin{equation}
t(p)= \exp(ipL) , \;\; r^{(l)}(p) = 0 , \;\; r^{(r)}(p)=\frac{i V_0 \Lambda}{2 \pi} \frac{\sin(\delta L)}{\delta} \exp[i(p+\pi/ \Lambda)L ] \;\;\;\;
\end{equation}
where we have set $\delta=p-\pi/\Lambda$.
Such equations clearly indicate that, for left-side incidence, the crystal appears to be fully invisible, i.e. there are not reflected waves and the transmitted ones propagate as if the crystal were absent \cite{L5}. Conversely, for right-side incidence as the transmitted wave propagates again as if the crystal were absent, a reflected wave is generated, with a reflectance $R^{(R)}=|r^{(l)}|^2$ that grows quadratically with the crystal thickness $L$ at Bragg resonance $\delta=0$, i.e. for $p=\pi/\Lambda$ . Such a physically relevant behavior was referred to as {\em unidirectional invisibility} in Ref.\cite{L5}.  As it is shown in the Appendix A and briefly mentioned in Ref.\cite{L5}, in the framework of the coupled-mode theory unidirectional invisibility is predicted to occur for a rather general class of complex potentials with zero mean, $V(x)= \sum_{n \neq 0} \Phi_n \exp(2 \pi i n x/ \Lambda)$, provided that the condition $\Phi_{-1}=0$ (or similarly $\Phi_{1}=0$) is satisfied. Note that the $\mathcal{PT}$-symmetric sinusoidal potential (2) at $\sigma=1$ belongs to such a general class of complex potentials.\\
 In the next section, we will derive {\em exact} expressions for transmission and reflection coefficients for the $\mathcal{PT}$-symmetric sinusoidal crystal, and will show that the invisibility property of the crystal, as predicted by Eq.(15), occurs solely for short crystals, and breaks down for long crystals. Here we discuss the reasons of failure of Eq.(15) to predict the correct expressions of transmission and reflection coefficients in long crystals, and propose an {\em extended} version of coupled-mode theory to properly describe Bragg scattering in complex crystals.
The derivation of coupled-mode equations is routinely done by means of averaging or multiple-scale asymptotic techniques, which is detailed in the Appendix A for the general case of a complex potential $V(x)=\sum_{n \neq 0} V_n \exp(2 i \pi n x / \Lambda)$ with zero mean. Here we give explicit analytical results for the sinusoidal $\mathcal{PT}$-symmetric crystal at $\sigma=1$, however as shown in the Appendix A similar results are obtained for a more general complex crystal provided that the condition $\Phi_{-1}=0$ is satisfied. For a small value of $\alpha$, a solution to Eq.(1) can be searched as a power series expansion
\begin{equation}
\psi(x)=\psi^{(0)}(x)+ \alpha \psi^{(1)}(x)+\alpha \psi^{(2)}(x)+...
\end{equation}
and multiple spatial scales $X_0=x$, $X_1= \alpha x$, ... are introduced to satisfy solvability conditions at the various orders in the asymptotic analysis.
If the analysis is pushed up to the order $\sim \alpha$, the solution to Eq.(1) inside the crystal for a small value of $\delta=p-\pi/\Lambda$ (of order $ \sim \alpha$) can be written as
\begin{equation}
\psi(x)=\psi^{(0)}(x)+ \alpha \psi^{(1)}(x)+o(\alpha^2)  
\end{equation}
where we have set 
\begin{eqnarray}
\psi^{(0)}(x) & = &  u(x) \exp(i \pi x/\Lambda)+v(x) \exp(-i \pi x / \Lambda)  \\
\alpha \psi^{(1)}(x) & = & \frac{V_0 \Lambda^2 u(x)}{8 \pi^2} \exp(3 i \pi x / \Lambda)
\end{eqnarray}
and where the amplitudes $u$ and $v$ satisfy the coupled-mode equations
\begin{eqnarray}
i \frac{du}{dx} & = & -\delta u-\frac{V_0 \Lambda}{2 \pi}v \\
i \frac{dv}{dx} & = &  \delta v
\end{eqnarray}
From Eqs.(20) and (21), it follows that the amplitudes $u$ and $v$ at the planes $x=0$ and $z=L$ are related by the relation $(u(L),v(L))^T=\mathcal{K}(p) (u(0),v(0))^T$, where the matrix $\mathcal{K}(p)$ reads explicitly
\begin{equation}
\mathcal{K}(p)= \left(
\begin{array}{cc}
\exp(i \delta L) & i \frac{V_0 \Lambda}{2 \pi} \frac{\sin (\delta L)}{\delta} \\
0 & \exp(-i \delta L)
\end{array}
\right).
\end{equation}
In standard coupled-mode theory \cite{Sipe,Poladian},  only the leading order term $\psi^{(0)}(x)$ in the expansion (17) is considered for the computation of the transfer matrix $\mathcal{M}$, and one simply has  
\begin{equation}
\mathcal{M}(p)=\mathcal{S}(p) \mathcal{K}(p) 
\end{equation}
where we have set
\begin{equation}
\mathcal{S}(p)=\left(
\begin{array}{cc}
\exp(i \pi L / \Lambda) & 0 \\
0 & \exp(-i \pi L / \Lambda)
\end{array}
\right).
\end{equation}
Using Eq.(23) for the expression of the transfer matrix, one then obtains Eq.(15) for the reflection and transmission coefficients. However, for a complex crystal such an analysis can fail even if the correction term $\alpha \psi^{(1)}$, given by Eq.(19),  remains smaller than $\psi^{(0)}$. To understand why this may happen, let us consider as an example the case of left-side incidence. This case requires the absence of incident waves  from the right side, which formally implies $(d \psi/dx)=ip \psi$ at $x=L$, where $p$ is the momentum of the incident wave from the left side.  If the approximation $\psi(x) \simeq \psi^{(0)}(x)$ is made, the appropriate boundary conditions would be $u(0)=1$ and $v(L)=0$. The key point is that, if the expression (17) of $\psi(x)$ up to the order $\sim \alpha$ is now considered, the boundary condition $v(L)=0$ does not exactly correspond to the absence of incident waves from the right side, just because of the (small) additional contribution to $\psi(x)$ given by $\alpha \psi^{(1)}(x)$ [Eq.(19)].  Hence a more accurate procedure should use the boundary conditions $u(0)=1$ and $v(L)=\epsilon$, where $\epsilon$ is a small parameter (of order $\alpha$) to be determined such that $(d \psi/dx)=ip \psi$ at $x=L$. It is obvious that, if the solution to the coupled-mode equations (20-21) with the boundary conditions $u(0)= 0$ and $v(0)=\epsilon$ (corresponding to probing the crystal from the {\em right} side with a small amplitude $\epsilon$) would remain small uniformly in the interval $(0,L)$, the additional contribution to the solution arising from taking $\epsilon \neq 0$ would just introduce a small correction to the solution corresponding to the boundary conditions $u(0)= 1$ and $v(0)=0$. Hence a small correction to the transmission and reflection coefficients Eq.(15) would be obtained. This case always occurs for short enough crystals. However, if the crystal length $L$ is long enough such that the reflection $r^{(r)}$ for right-side incidence becomes large (of the order or larger than $ \sim 1 / \alpha$), then the correction arising from the solution  to the coupled-mode equations (20-21) with the boundary conditions $u(0)= 0$ and $v(0)=\epsilon$ can not be neglected anymore, and thus should be accounted for when calculating the reflection and transmission coefficients with the appropriate boundary conditions. The crystal length $L$ at which failure of Eq.(15) is expected to occur can be estimated by imposing $|r^{(r)}| \simeq 1 / \alpha$. Taking for $|r^{(r)}|$ its peak value at $\delta=0$, i.e. $|r^{(r)}| \sim V_0 \Lambda L /(2 \pi)$, one then expects failure of Eq.(15) for $L \geq \sim L_c$, where
\begin{equation}
L_c =  \frac{2 \pi^3}{V_0^2 \Lambda^3}. 
\end{equation}
Hence, for $L$ of the order of larger than $L_c$, Eq.(15) can not be used to calculate the transmission and reflection coefficients of the crystal. A more appropriate procedure to compute the transfer matrix, and thus the transmission and reflection coefficients, is to use Eq.(10), in which the fundamental matrix $\mathcal{Z}$ is  calculated using for $\psi(x)$ the expression given by Eqs.(17-19), i.e. {\em including} the first-order correction term $\alpha \psi^{(1)}(x)$. Such an extension of the ordinary coupled-mode theory will be referred to as the {\it extended} coupled-mode theory.

\section{Bragg scattering in sinusoidal $\mathcal{PT}$-symmetric crystals: Exact analysis}

Exact expressions for the reflection and transmission coefficients can be derived for the sinusoidal potential (2) at $\sigma=1$ in terms of modified Bessel functions of firts kind. In fact, after the change of variable $y= ( \Lambda \sqrt{V_0}/ \pi) \exp(i \pi x / \Lambda)$, Eq.(1) reduces to the Bessel equation \cite{L4}
\begin{equation}
y^2 \frac{d^2 \psi}{dy^2}+y \frac{d \psi}{dy}-(y^2+q^2) \psi=0
\end{equation}
where we have set
\begin{equation}
q= \frac{p \Lambda}{\pi}.
\end{equation}
Note that Bragg scattering for particles with momentum $p$ close to $ \pi / \Lambda$ corresponds to $q \sim 1$. For $ q \neq 1$, two linearly independent solutions to Eq.(26) are $I_q(y)$ and $I_{-q}(y)$, where $I_q(y)$ is the modified Bessel function $I$ of first kind   \cite{Abramowitz}. Hence in the interval $0<x<L$ two linearly independent solutions to Eq.(1) are given by
\begin{equation}
\Phi_1(x)=I_q(\Delta \exp(i \pi x/\Lambda)) \; , \;\;  \Phi_2(x)=I_{-q}(\Delta \exp(i \pi x/\Lambda)) 
\end{equation}
where we have set
\begin{equation}
\Delta=\frac{\Lambda \sqrt{V_0}}{ \pi}= \sqrt{\alpha}.
\end{equation}
Using Eq.(28), one can construct the fundamental matrix ${\mathcal Z}(x)$ of Eq.(1) from $x=0$ to $x=L$ as
\begin{equation}
\mathcal{Z}(p) = \left(
\begin{array}{cc}
\Phi_1(L) & \Phi_2(L) \\
\Phi_1^{'}(L) & \Phi_2^{'}(L)
\end{array}
\right) \times 
\left(
\begin{array}{cc}
\Phi_1(0) & \Phi_2(0) \\
\Phi_1^{'}(0) & \Phi_2^{'}(0)
\end{array}
\right) ^{-1}
\end{equation}
where the apex denotes the derivative with respect to $x$. The transfer matrix $\mathcal{M}(p)$ is finally obtained after substitution of Eq.(30) into Eq.(10). After some lengthy calculations, which are briefly detailed in the Appendix B, the following expressions for the transfer matrix coefficients are obtained:
\begin{eqnarray}
\mathcal{M}_{11}(p)  =  \cos(pL)+ i \frac{\Lambda \sin (pL) }{2p \sin( \pi q)} \left( p^2 Q_1 Q_2- V_0 D_1 D_2  \right) \\
\mathcal{M}_{12}(p)  = -  i \frac{\Lambda \sin (pL) }{2p \sin( \pi q)} \left[ V_0 D_1 D_2 +p^2 Q_1 Q_2 +p \sqrt{ V_0} \left( D_1Q_2+D_2Q_1 \right)   \right] \\
\mathcal{M}_{21}(p)  =  i \frac{\Lambda \sin (pL) }{2p \sin( \pi q)} \left[ V_0 D_1 D_2 +p^2 Q_1 Q_2 -p \sqrt{V_0} \left( D_1Q_2+D_2Q_1 \right)   \right] \\
\mathcal{M}_{22}(p)  =  \cos(pL)- i \frac{\Lambda \sin (pL) }{2p \sin( \pi q)} \left( p^2 Q_1 Q_2- V_0 D_1 D_2  \right) 
\end{eqnarray}
where we have set
\begin{equation}
Q_1=I_q(\Delta) \; , \; Q_2=I_{-q}(\Delta) \; , \; D_1=I^{'}_q(\Delta) \; , \; D_2=I_{-q}^{'}(\Delta)
\end{equation}
and where $q$ and $\Delta$ are defined by Eqs.(27) and (29), respectively. The reflection and transmission coefficients $r^{(l,r)}(p)$ and $t(p)$ are then obtained after substitution of Eqs.(31-34) into Eq.(8). In particular, for the transmission coefficient $t(p)$ one  obtains explicitly
\begin{equation}
t(p)=\frac{1}{\cos(pL)-iF(p) \sin(pL)}
\end{equation}
where we have set
\begin{equation}
F(p)=\frac{\Lambda}{2p \sin( \pi q)} \left(p^2Q_1Q_2-V_0D_1D_2 \right).
\end{equation}
Note that Eq.(36) would reduce to the first of the Eq.(15), corresponding to unidirectional crystal invisibility, if the function $F(p)$ were replaced by $1$.\\
The previous equations (31-37) have been derived for Bragg scattering of matter waves in the framework of the Schr\"{o}dinger equation (1), however similar relations hold for Bragg scattering of light waves in a complex Bragg grating structure, governed by the similar equation (3). Specifically, in view of Eqs.(4) and (5), for a grating structure with a sinusoidal $\mathcal{PT}$-symmetric modulation of the complex relative dielectric constant  $\Delta \epsilon(x)= \Phi \exp(2 i \pi x / \Lambda)$ and for incident light waves with frequency $\omega$,  the analytical expressions given above still hold, provided that the following formal substitutions $ p \rightarrow \omega n_0 /c_0$ and $V_0 \rightarrow (n_0 \omega/c_0)^2 \Phi \simeq (\pi/\Lambda)^2 \Phi$ are made, where $n_0$ is the refractive index of the lossless dielectric medium and $c_0$ the speed of light in vacuum.

\section{Unidirectional crystal invisibility}
Let us know discuss the unidirectional invisibility of the sinusoidal $\mathcal{PT}$-symmetric crystal on the basis of the exact scattering results presented in the previous section. To study the exact behavior of $t(p)$ as given by Eq.(36) and breakdown of crystal transparency as the number of cells $N$ is increased,  let us measure the length $x$ in units of $\Lambda / \pi$, i.e. let us set without loss of generality $\Lambda=\pi$. With such a scaling, one has $\alpha=V_0$, $q=p$, $L= N \pi$ and $\Delta= \sqrt{V_0}$. Using the identity $I^{'}_p (\Delta)=I_{p-1}(\Delta)-(p/\Delta) I_{p}(\Delta)=I_{p+1}(\Delta)+(p/\Delta) I_{p}(\Delta)$ for the derivative of modified Bessel functions \cite{Abramowitz}, one can write
\begin{equation}
t(p)=\frac{1}{\cos(N \pi p)-i
F(p) \sin(N \pi p)}
\end{equation}
with
\begin{equation}
F(p)=\frac{\pi}{2p \sin( \pi p)} \left[ \sqrt{V_0} p (I_{-p} I_{p-1}+ I_pI_{-p+1})-V_0 I_{p-1}I_{-p+1}  \right]
\end{equation}
\begin{figure}[htb]
\centerline{\includegraphics[width=10cm]{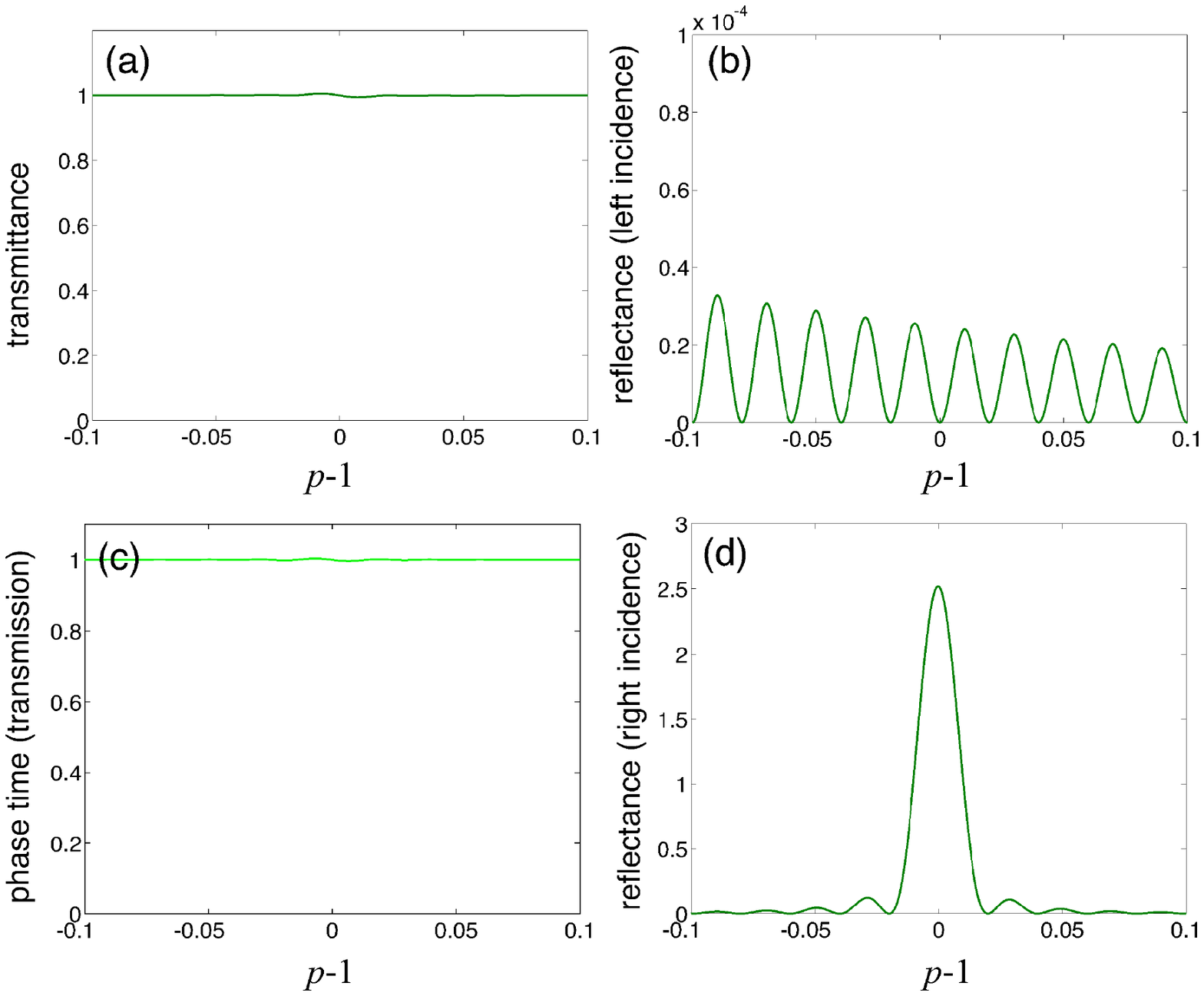}} \caption{
Behavior of (a) transmittance, (b) reflectance for left-side incidence, (c) normalized phase time (in transmission), and (d) reflectance for right-side incidence versus momentum $p$ of the incident particle in a sinusoidal $\mathcal{PT}$-symmetric crystal for parameter values $\Lambda= \pi$, $V_0=0.02$ and for a number of cells $N=50$. The curves are obtained by using the exact expressions Eqs.(31-34)  for the transfer matrix coefficients.}
\end{figure}
\begin{figure}[htb]
\centerline{\includegraphics[width=10cm]{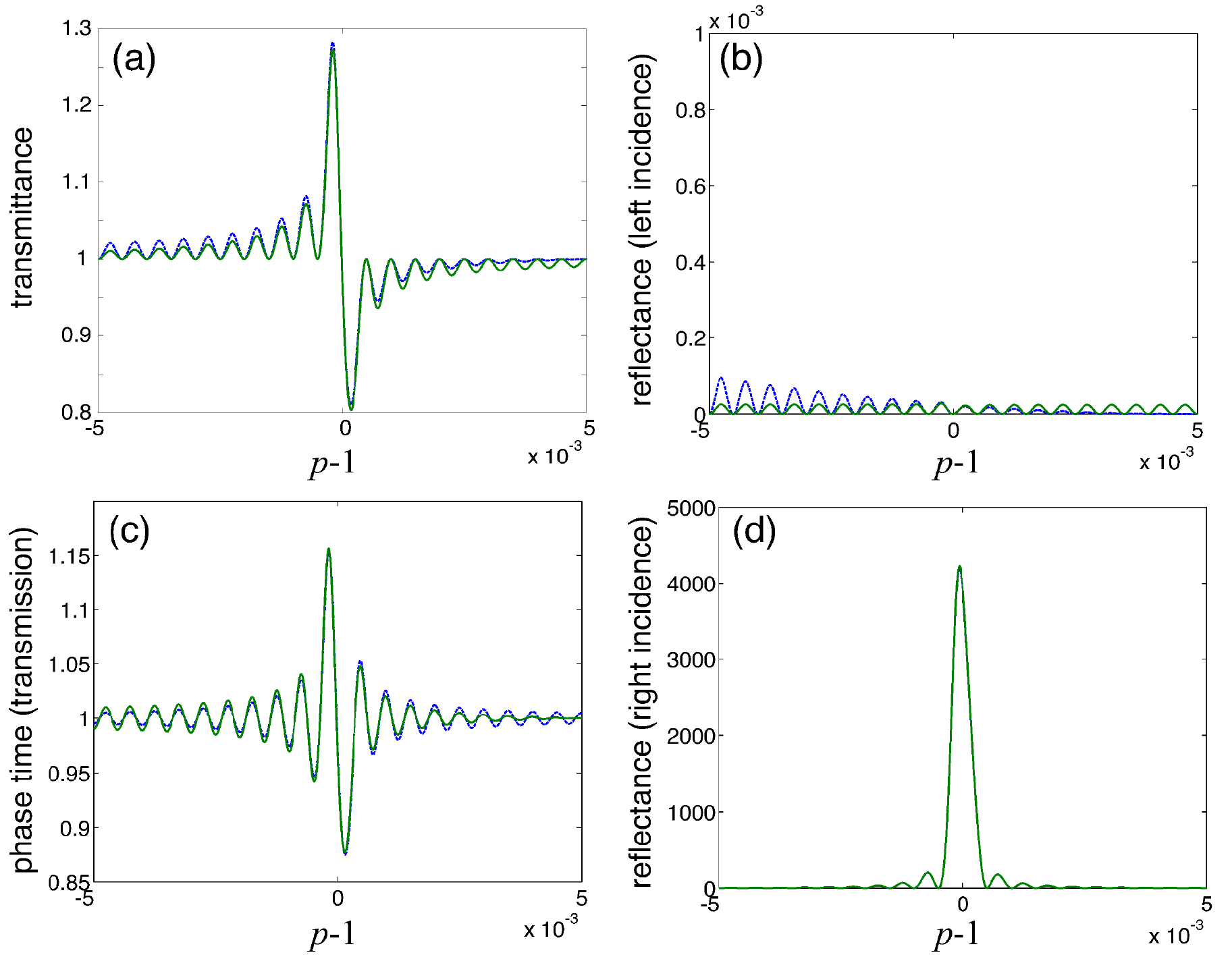}} \caption{
 Same as Fig.1, but for $N=2000$. In the figures, solid lines refer to the exact scattering results obtained from Eqs.(31-34), whereas the dashed curves are the predictions based on the extended coupled-mode theory.}
\end{figure}
and where the Bessel functions are calculated at $\sqrt{V_0}$ . Note that $t(p)$ depends on two parameters solely: the number of crystal cells $N$ and the potential amplitude $V_0$. Note also that Eq.(38) is valid regardless of the smallness of $V_0$ and far from the Bragg resonance condition $p=1$ as well.  Similar expressions can be derived for the reflection coefficients $r^{(l,r)}(p)$ in terms of modified Bessel functions. In the computation of the reflection and transmission coefficients, the modified Bessel functions have been calculated using a fast and highly accurate routine, discussed in Ref. \cite{Amos}. The accuracy of our procedure has been tested by checking the agreement of the results obtained from the exact analytical prediction [Eqs. (38) and (39)] and from the full numerical procedure outlined in Sec.2.2 [Eqs.(10-14)]. \par 
Figures 1 and 2 show the behaviors of spectral transmittance $T(p)=|t(p)|^2$ and reflectances $R^{(l,r)}(p)=|r^{(l,r)}(p)|^2$ for left and right side incidence, as calculated by the exact analysis (solid curves), for increasing values of the number of cells $N$ and for $V_0=0.02$. In the figures, the behavior of the normalized phase time of transmitted waves, defined by $\tau_t(p)=(1/L)(d \phi_t / dp)$, is also depicted, where $\phi_t(p)$ is the phase of $t(p)$. Physically, $\tau_t(p)$ represents the traversal time of a narrow wave packet, with central momentum $p$, across the crystal, normalized to the transit time in vacuum (i.e. in the absence of the crystal). For a relatively small numbers of cells, as in Fig.1, unidirectional invisibility is observed, and reflection for left-side crystal incidence is extremely small, according to the coupled-mode theory of Sec.2.3 and the results of Refs.\cite{L5,cg}. However, as the number of cells is increased to become comparable or larger than $N_c=L_c/ \Lambda$, given by   [see Eq.(25)]
\begin{equation}
N_c \sim \frac{L_c}{\Lambda}=\frac{2}{\pi \alpha^2},
\end{equation}
the invisibility regime breaks down near $p=1$, with the appearance of oscillations of the transmittance and phase time, as one can clearly see in Figs.2(a) and (c). In the figures, the predictions of spectral transmittance and reflectance computed by the {\em extended} coupled-mode theory, discussed at the end of Sec.2.3, are also depicted by the dotted curves.  Note that, as the crystal is not anymore transparent, the reflectance for left-side incidence remains extremely small [see Fig. 2(b)]. In such a regime the crystal is not invisible, however it is still unidirectional {\em reflectionless}. As the number of cells is further increased, the transmittance grows in a narrow interval near the Bragg condition $p \simeq 1$, as well the the reflectance for right-side incidence. As the transmittance becomes large enough, such that $\mathcal{M}_{22}$ becomes smaller and of the same order of magnitude than $\mathcal{M}_{21}$,  a very narrow resonance peak appears in the reflectance spectrum for left-side incidence. This is shown in Fig.3, in which solid and dotted curves refer to the exact results and to the approximate ones based on the extended coupled-mode theory, respectively. In such a regime, both unidirectional invisibility and reflectionless properties of the complex crystal are thus broken.To estimate the number of cells $N_c^{'}>N_c$ at which such a second transition occurs, we can apply the extended coupled-mode theory, discussed in Sec.2.3, to calculate an approximate expression of the reflection coefficient $r^{(l)}$ for left-side incidence. Using Eqs.(17-21) for an approximate expression of $\psi(x)$ and applying the appropriate boundary conditions, corresponding to left-side incidence, at exact Bragg resonance ($p=\pi/\Lambda$)  an approximate expression for $|r^{(l)}|$ can be derived, which reads explicitly $r^{(l)} \sim (\pi / 64) \alpha^3 (L/ \Lambda)$. The critical number of cells $N_{c}^{'}$ at which the crystal is not anymore reflectionless for left-side incidence can be estimated by letting $|r^{(l)}| \sim 1$, which yields
\begin{equation}
N_c^{'} \sim \frac{64}{\pi \alpha^3} .
\end{equation}

\begin{figure}[htb]
\centerline{\includegraphics[width=10cm]{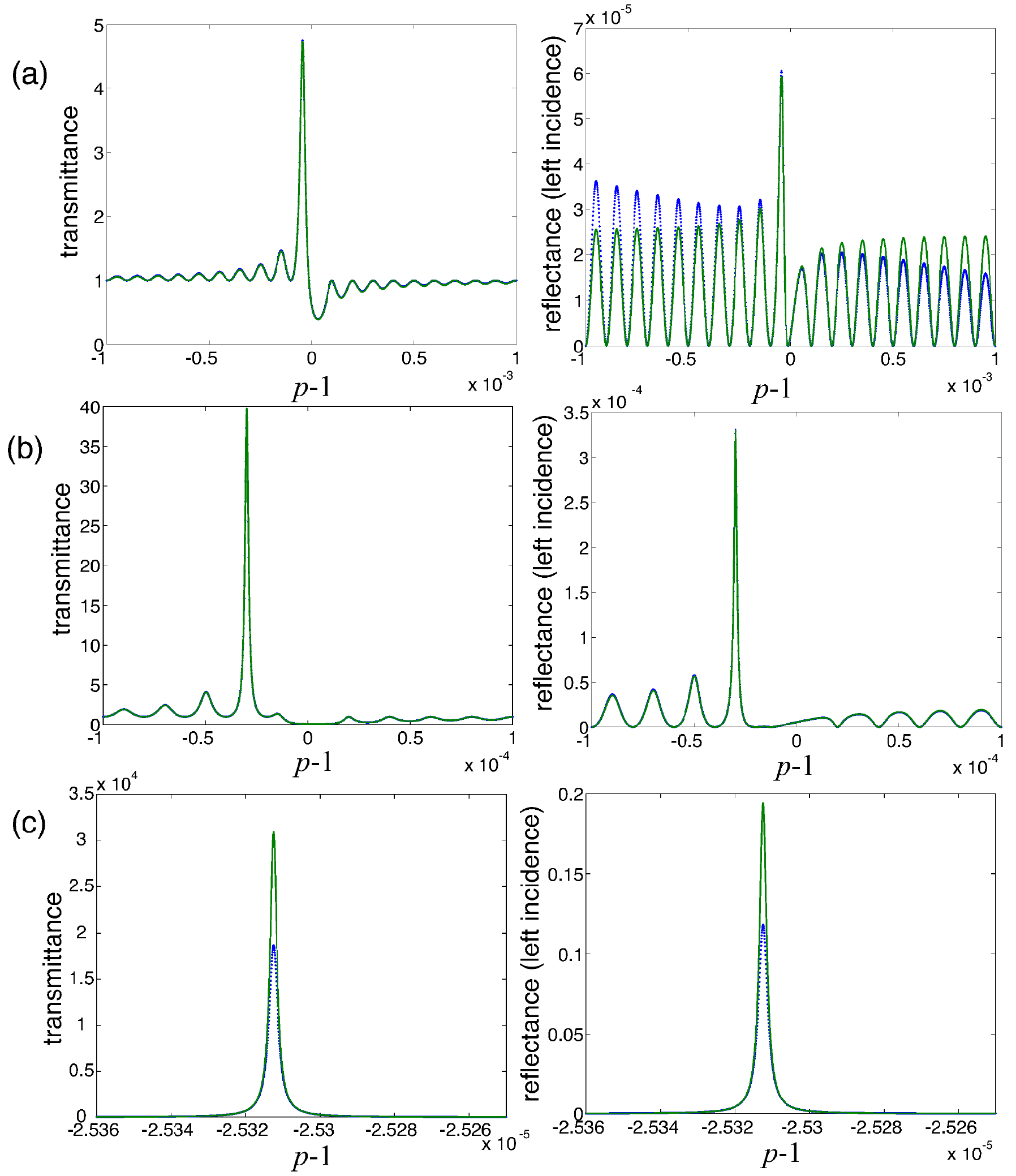}} \caption{
 Transmittance (left panels) and reflectance for left-side incidence (right panels) in a sinusoidal $\mathcal{PT}$-symmetric crystal for $V_0=0.02$, $\Lambda= \pi$ and for increasing number of crystal cells: (a) $N=10000$, (b), $50000$, and (c) $N=1600000$.}
\end{figure}
\section{Conclusion}
In this work Bragg scattering in sinusoidal $\mathcal{PT}$-symmetric complex crystals of finite thickness has been theoretically investigated, and exact analytical expressions for reflection and transmission coefficients have been derived in terms of 
modified Bessel functions. The analytical results indicate that unidirectional invisibility, recently predicted for such crystals  by coupled-mode theory \cite{L5,cg},
breaks down for crystals containing a large number of unit cells. In particular, for a given modulation depth in a shallow sinusoidal potential, three regimes have been found as the crystal length is increased. At short lengths  the crystal is reflectionless and invisible when probed from one side (unidirectional invisibility), according to standard coupled-mode theory. As the numbers of cells is increased, absence of reflection for one side incidence is still observed, however the crystal is no more invisible because large oscillations in the transmittance and in the transmission phase time appear near the Bragg resonance. For still thicker crystals, both unidirectional reflectionless and invisibility properties are broken. 
An extension of coupled-mode theory has been proposed to properly modelling the scattering properties of complex crystals.

\appendix
\section{Derivation of coupled-mode equations}

In this Appendix we briefly derive coupled-mode equations describing first-order Bragg scattering in a complex crystal with a shallow lattice in the framework of the Schr\"{o}dinger equation (1).
Let us consider Bragg scattering of a particle with momentum $p$ close to the Bragg value $\pi/ \Lambda$, i.e. with energy $E=p^2$ close to $(\pi/\Lambda)^2$, and let us assume for the complex scattering potential $V(x)$ a rather general profile with zero mean, given by the Fourier expansion 
\begin{equation}
V(x)= \sum_{n \neq 0} \Phi_n \exp(2 i \pi n x / \Lambda)
\end{equation}
for $0<x<L$. The shallow lattice approximation implies that the Fourier amplitudes $\Phi_n$ are much smaller than $E$.
Note that the $\mathcal{PT}$-symmetric sinusoidal potential (2) is obtained as a special case of Eq.(A.1) after setting $\Phi_1=(V_0/2)(1 + \sigma)$, $\Phi_{-1}=(V_0/2) (1-\sigma)$ and $\Phi_n=0$ for $n \neq  \pm1$.  To develop a perturbative analysis of Eq.(1) in the shallow grating approximation $V(x) \rightarrow 0$ and for $p \rightarrow \pi/ \Lambda$, it is worth introducing a parameter $\alpha$ that measures the smallness of the various terms entering in the equations and rewriting Eq.(1) in the following form, suited for an asymptotic analysis
\begin{equation}
\frac{d^2 \psi}{dx^2}+\left( \frac{\pi}{\Lambda}\right)^2 \psi= -\alpha \left[ V(x) \psi+ W \psi \right],
\end{equation}
where we have set 
\begin{equation}
W \equiv E-\left( \frac{\pi}{\Lambda} \right)^2 \simeq \frac{2 \pi}{\Lambda} \left( p-\frac{\pi}{\Lambda}  \right) . 
\end{equation}
The problem is to construct
an asymptotic approximation of the perturbed solution
$\psi=\psi(x; \alpha)$ to Eq.(A.2) as $\alpha \rightarrow 0$. Therefore,
we seek a perturbation expansion of $\psi$ in the form of a power series in $\alpha$
\begin{equation}
\psi(x; \alpha)=\psi^{(0)}(x)+ \alpha \psi^{(1)}(x)+ \alpha^2 \psi^{(2)}(x)+ ...
\end{equation}
and introduce multiple scales for space $x$
\begin{equation}
X_0=x \; , \; \; X_1= \alpha x \; , \; \;, X_2= \alpha^2 x \; , ....
\end{equation}
which are needed to satisfy the solvability conditions in the asymptotic expansion at various orders. 
Substitution of the Ansatz (A.4) into Eq.(A.2) and using the derivative rule
  \begin{equation}
 \frac{d^2}{d x^2}= \frac{\partial^2}{\partial X_0^2}+2 \alpha \frac{\partial}{\partial X_0} \frac{ \partial}{\partial X_1}+ \alpha^2 \left( \frac{\partial^2}{\partial X_1^2}+2 \frac{ \partial}{\partial X_0} \frac{\partial}{\partial X_2} \right)+...
 \end{equation}
yields a hierarchy of
equations for successive corrections to $\psi$, which are obtained 
after collecting the terms of
the same order in $\alpha$ in the equation so obtained. At leading order $\sim \alpha^{0}$ one has
\begin{equation}
\frac{\partial ^2 \psi^{(0)}}{\partial X_0^2}+\left( \frac{\pi}{\Lambda}\right)^2 \psi^{(0)}=0
\end{equation}
whose general solution is given by
\begin{eqnarray}
\psi^{(0)}(X_0,X_1,X_2,...) & = & u(X_1,X_2,...) \exp(i \pi X_0 / \Lambda)+ \nonumber \\
& + & v(X_1,X_2,...) \exp(-i \pi X_0 / \Lambda) \; \; \; \;\;\;\;\;\;
\end{eqnarray}
where the amplitudes $u$ and $v$ may vary over the slow spatial scales $X_1$, $X_2$, ...  At order $\sim \alpha$, one obtains
\begin{equation}
\frac{\partial ^2 \psi^{(1)}}{\partial X_0^2}+\left( \frac{\pi}{\Lambda}\right)^2 \psi^{(1)}=g^{(1)}(X_0)
\end{equation}
where the forcing term $g^{(1)}$ is given by
\begin{equation}
g^{(1)}(X_0)=-\left[ V(X_0)+W \right] \psi^{(0)}-2 \frac{\partial^2 \psi^{(0)}}{\partial X_0 \partial X_1}.
\end{equation}
The solvability condition for Eq.(A9) requires that the forcing term $g^{(1)} (X_0)$ does not contain terms oscillating like $ \sim \exp( \pm i \pi X_0 / \Lambda)$. 
After substitution of Eqs.(A.1) and (A.8) into Eq.(A.10) and letting equal to zero the coefficients of the terms oscillating like $ \sim \exp(\pm i \pi X_0 / \Lambda)$ in the expression so obtained, the following equations for the 
evolution of the amplitudes $u$ and $v$ on the slow spatial scale $X_1$ are then obtained
\begin{eqnarray}
2 i \frac{\pi}{\Lambda} \frac{\partial u}{\partial X_1} & = & -Wu-\Phi_1v \\
2 i \frac{\pi}{\Lambda} \frac{\partial v}{\partial X_1} & = & Wv+\Phi_{-1} u
\end{eqnarray}
and the solution at order $ \sim \alpha$ is given by
\begin{eqnarray}
\psi^{(1)}(X_0) & = & \frac {\Lambda^2}{\pi^2} u \sum_{n \neq 0,-1} \frac{\Phi_n \exp[i \pi (2n+1) X_0 / \Lambda]}{(2n+1)^2-1}+ \nonumber \\
& + & \frac {\Lambda^2}{\pi^2} v \sum_{n \neq 0,1} \frac{\Phi_n \exp[i \pi (2n-1) X_0 / \Lambda]}{(2n-1)^2-1}
\end{eqnarray}
The evolution equations of the envelopes $u$ and $v$ at longer spatial scales $X_2$, $X_3$, ... are obtained similarly as solvability conditions at orders $\alpha^2$, $\alpha^3$, .... in the asymptotic expansion. The evolution of the amplitudes $u$ and $v$ in the physical spatial variable $x$ are then given by $du/dx = \alpha \partial_{X_1} u+ \alpha^2 \partial^2_{X_2} u + ....$ and $dv/dx= \alpha \partial_{X_1} v+ \alpha^2 \partial^2_{X_2} v + ....$. If we limit our analysis to the order $\sim \alpha$, after setting $\alpha=1$ from Eqs.(A.11) and (A.12) and using Eq.(A.3) one finally obtains the following coupled-mode equations for the envelopes $u$ and $v$
\begin{eqnarray}
i\frac{du}{dx} & = & -\delta u-\rho_1 v \\
i\frac{dv}{dx} & = & \delta u+\rho_2 u
\end{eqnarray}
where we have set
\begin{equation}
\delta= p-\frac{\pi}{\Lambda} \; ,\;\; \rho_1= \frac{ \Lambda \Phi_1}{ 2 \pi} \; , \;\; \rho_2= \frac{ \Lambda \Phi_{-1}}{ 2 \pi}.
\end{equation}
In the framework of the standard coupled-mode theory \cite{L5,cg}, unidirectional crystal invisibility is attained whenever the evolution of either one of the two amplitudes $u$ or $v$ is decoupled from the other one, i.e. for either $\Phi_{-1}=0$ or $\Phi_1=0$.  In particular, for the $\mathcal{PT}$-symmetric crystal [Eq.(2)] at $\sigma=1$, one has $\Phi_1=V_0$ and $\Phi_{-1}=0$, which ensures unidirectional invisibility. In this case, the coupled-mode equations (A.14) and (A.15), as well as the first-order correction $\psi^{(1)}$ as given by Eq.(A.13), reduce to Eqs.(19), (20) and (21) given in the text.

\section{Derivation of the transfer matrix}
The exact expression of the transfer matrix $\mathcal{M}(p)$ is obtained using Eq.(10), where the fundamental matrix $\mathcal{Z}(p)$ is calculated in terms of modified Bessel functions according to Eq.(30). A simplified form of $\mathcal{Z}(p)$ can be obtained after observing that, owing to the analytic continuation of the Bessel $I$ function in the complex plane \cite{Abramowitz}, one has 
\begin{equation}
\Phi_1(x+L)=\Phi_1(x) \exp(i \pi q L / \Lambda) , \;\;   \Phi_2(x+L)=\Phi_2(x) \exp(-i \pi q L / \Lambda) \;\;\;
\end{equation}
and thus $\Phi_{1,2}(L)=\Phi_{1,2}(0) \exp( \pm i \pi q L / \Lambda)$ and $\Phi^{'}_{1,2}(L)= \Phi^{'}_{1,2}(0) \exp(\pm i \pi q L / \Lambda)$. Hence one can write
\begin{equation}
\left(
\begin{array}{cc}
\Phi_1(L) & \Phi_2(L) \\
\Phi_1^{'}(L) & \Phi_2^{'}(L)
\end{array}
\right)=  
\left(
\begin{array}{cc}
\Phi_1(0)  & \Phi_2(0) \\
\Phi_1^{'}(0) & \Phi_2^{'}(0)
\end{array}
\right)
\left(
\begin{array}{cc}
\exp(i \pi q L / \Lambda)  & 0\\
0 &\exp(-i \pi q L / \Lambda)
\end{array}
\right). \;\;\;
\end{equation}
Moreover, taking into account the property of the Wronskian of modified Bessel functions \cite{Abramowitz}, one has
\begin{equation}
\left|
\begin{array}{cc}
\Phi_1(0) & \Phi_2(0) \\
\Phi_1^{'}(0) & \Phi_2^{'}(0)
\end{array}
\right|=i \frac{\pi \Delta}{\Lambda}\left|
\begin{array}{cc}
I_q(\Delta) & I_{-q}(\Delta) \\
I_{q}^{'}(\Delta) & I_{-q}^{'}(\Delta)
\end{array}
\right|=-i\frac{2}{\Lambda} \sin( q \pi)
\end{equation}
and hence
\begin{equation}
\left(
\begin{array}{cc}
\Phi_1(0)  & \Phi_2(0) \\
\Phi_1^{'}(0) & \Phi_2^{'}(0)
\end{array}
\right)^{-1}=i \frac{\Lambda}{2 \sin (q \pi)} \left(
\begin{array}{cc}
\Phi_2^{'}(0)  & -\Phi_2(0) \\
-\Phi_1^{'}(0) & \Phi_1(0)
\end{array}
\right).
\end{equation}
Substitution of Eqs.(B.2) and (B.4) into Eq.(30) yields for the fundamental matrix $\mathcal{Z}(p)$ the following simplified expression
\begin{eqnarray}
\mathcal{Z}(p)=i \frac{\Lambda}{2 \sin (q \pi)} \left(
\begin{array}{cc}
\Phi_1(0)  & \Phi_2(0) \\
\Phi_1^{'}(0) & \Phi_2^{'}(0)
\end{array}
\right)
\left(
\begin{array}{cc}
\exp(i \pi q L / \Lambda)  & 0\\
0 &\exp(-i \pi q L / \Lambda)
\end{array}
\right) \times \nonumber \\
\times \left(
\begin{array}{cc}
\Phi_2^{'}(0)  & -\Phi_2(0) \\
-\Phi_1^{'}(0) & \Phi_1(0)
\end{array}
\right)
\end{eqnarray}
where $\Phi_1(0)=Q_1$, $\Phi_2(0)=Q_2$, $\Phi_{1}^{'}(0)=i  \sqrt{ V_0 }D_1$, $\Phi_{2}^{'}(0)=i  \sqrt{ V_0 } D_2$ and $Q_{1,2}$, $D_{1,2}$ are defined by Eq.(35) given in the text. Substitution of Eq.(B.5) into Eq.(10) and using the expression of the matrix $\mathcal{T}(p)$ given by Eq.(11), after some lengthy but straightforward calculations one finally obtains for the coefficients of the transfer matrix $\mathcal{M}(p)$ the expressions given by Eqs.(31-34).

\end{document}